\documentclass[12pt,preprint]{aastex}
\usepackage{natbib}
\bibpunct{(}{)}{;}{a}{}{,}
\shorttitle{FUV diffuse emission from the LMC}
\shortauthors{Pradhan, Pathak \& Murthy}

\begin{document}

\title{Far Ultraviolet Diffuse Emission from the Large Magellanic Cloud}
\author{Ananta C. Pradhan, Amit Pathak and Jayant Murthy}
\affil{Indian Institute of Astrophysics, Koramangala, Bangalore-560034, India}

\begin{abstract}
We present the first observations of diffuse radiation in the far ultraviolet (1000 -- 1150 \AA) from the Large Magellanic Cloud based on observations made with the {\it Far Ultraviolet Spectroscopic Explorer}. The fraction of the total radiation in the field emitted as diffuse radiation is typically 5 -- 20\%  with a high of 45\% near N70 where there are few exciting stars, indicating that much of the emission is not due to nearby stars. Much less light is scattered in the far ultraviolet than at longer wavelengths with the stellar radiation going into heating the interstellar dust.
\end{abstract} 

\section{Introduction}

The Large Magellanic Cloud (LMC) provides a very different view of the interstellar medium (ISM) than is possible either from observations of the Milky Way (MW) or of other, more distant galaxies. At a distance of 50 kpc \citep{Feast99}, it is far enough that we can get an overview of processes in the entire galaxy but close enough that we can distinguish between the different sources, particularly because of the low inclination angle of $35^{\circ}$ \citep{van der Marel01}. The most comprehensive survey of dust and gas in the LMC has come from \citet{Bernard08} using a variety of sources including new observations with the {\it Spitzer Space Telescope}. They found that there was a modification of the dust population in the LMC as compared with the Milky Way, perhaps through the erosion of large grains in the ISM.

The first imaging observations of the LMC in the ultraviolet (UV) were by \citet{Page81} using the S201 far ultraviolet (FUV) camera during the {\it Apollo 16} mission followed by \citet{Smith87} from a rocket experiment and \citet{Parker98} from the {\it Ultraviolet Imaging Telescope (UIT)} on board the Space Shuttle {\it Columbia}. Observations of the diffuse UV light track the transfer of radiation from the stellar radiation to the interstellar medium with the absorbed radiation re-emitted as thermal emission in the infrared. The diffuse UV radiation is about 25\% of the total radiation emitted from the LMC \citep{Parker98} and understanding its distribution is important to models of galactic evolution. More recently, \citet{Cole99a} used the rocket-borne {\it Wide-Field Imaging Survey Polarimeter (WISP)} to map the scattered light in the near ultraviolet (2150 \AA), finding that the scattered light is actually a complex combination of the relative geometry of the dust and the stars. It is not sufficient to merely have bright stars or to have dust: both must be present to show the scattered light. In this work, we use serendipitous observations made with the {\it Far Ultraviolet Spectroscopic Explorer (FUSE)} to report, for the first time, measurements of the diffuse FUV (1000 - 1150 \AA) emission in an external galaxy.

\section{Observations and Data Analysis}

\begin{figure}
\plotone{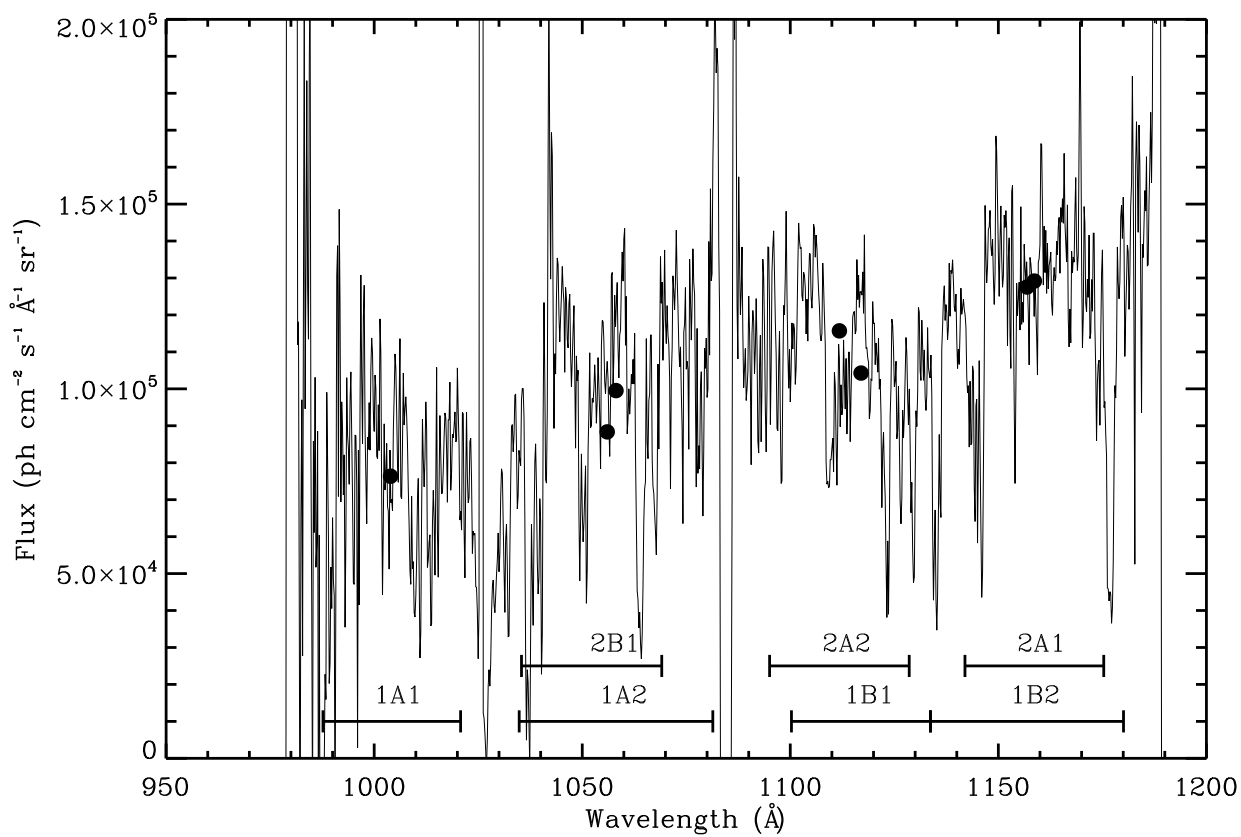}
\caption{Spectrum of N11 (PGMW-3053 in Table \ref{tbl-1}) showing the seven wavelength bands that have been integrated to obtain the diffuse emission. The wavelength range is shown by horizontal bars at the bottom of the spectrum. The integrated flux of each band is marked by solid circle at their effective wavelengths.
\label{Fig1}}
\end{figure}

The {\it FUSE} spacecraft and its mission have been described by \citet{Moos00} and \citet{Sahnow00}. The primary purpose of the mission was to take high resolution spectra ($\lambda/\Delta \lambda$ $\approx$ 20,000) of galactic and extragalactic sources. {\it FUSE}~ included 3 apertures: the HIRS (\(1.25''\) x \(20''\)); the MDRS (\(4''\) x \(20''\)); and the LWRS \((30''\) x \(30''\)), all of which obtained data simultaneously. Although only the LWRS with its relatively large field of view was useful for diffuse observations, there were many fields in which the primary aperture was either the HIRS or the MDRS leaving the LWRS aperture (separated from the other two apertures by \(100''\) and \(200''\), respectively) to observe a nominally blank region of the sky. \citet{Murthy04} have described the analysis of these serendipitous background observations and we have followed their extraction of diffuse surface brightnesses from the {\it FUSE} spectra, except that we have used the latest version of the data pipeline software (CalFUSE v3.2; \citet{Dixon07}). This involved treating {\it FUSE} as a broad band photometer and collapsing the spectra into two wavelength bands per detector, excluding the terrestrial airglow lines (primarily Ly$\beta$). This resulted in seven wavelength bands with effective wavelengths of 1004 \AA~(1A1), 1058 \AA~(1A2), 1117 \AA~(1B1), 1157 \AA~(1B2), 1159 \AA~(2A1), 1112 \AA~(2A2), 1056 \AA~(2B1). The instrumental background was derived from strips off the spectrum and subtracted from the band fluxes. Our wavelength bands are shown in Fig. \ref{Fig1} on top of a spectrum of the diffuse emission in N11, one of the brighter diffuse regions of the LMC. The integrated flux of each band is marked by a solid circle at its effective wavelength.

There are more than 600 {\it FUSE} pointings in and around the LMC and we have examined all for suitability for diffuse measurements. We immediately rejected all observations which specifically observed a bright star in the LWRS aperture but there were others where a star was coincidentally in the aperture. These were identified and rejected through their FWHM, which is less than 20 pixels for a point source but about 30 pixels for an aperture filling diffuse source. We were finally left with 81 observations of the diffuse radiation in different parts of the LMC.

These observations are from two classes of targets: observations of point sources through either the MDRS or HIRS apertures; or calibrations where the apertures were pointed at nominally blank areas of the sky in order to allow the spectrographs to thermalize before an instrumental realignment. As a result, our locations were generally around popular targets in the LMC such as 30 Doradus and N11 (Fig. \ref{Fig2}). These are tabulated in Table \ref{tbl-1} (see the online version of the Journal for the complete table).

\section {Results and Discussion} 

\begin{figure}
\plotone{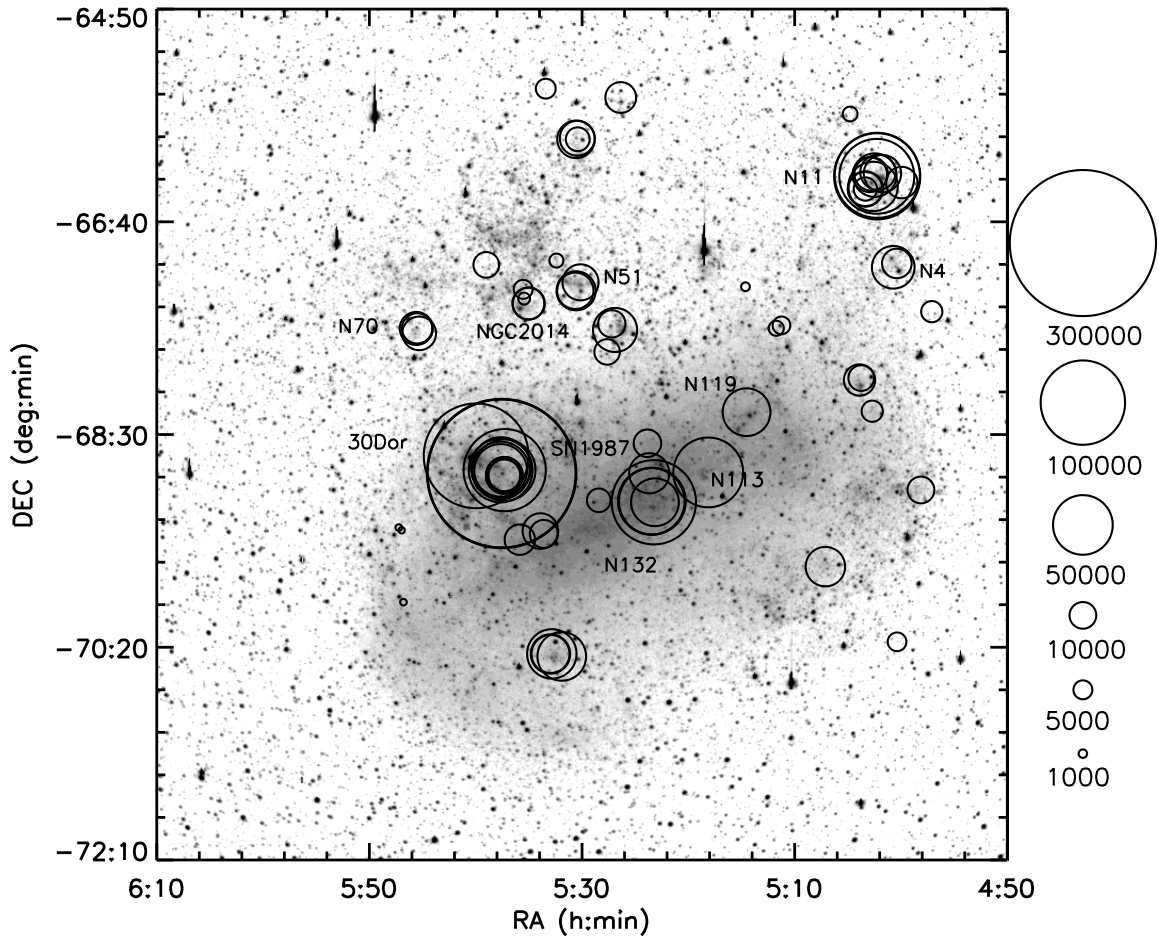}
\caption{LMC R-band image from \citet{Bothun88} with the {\it FUSE} observations represented as circles with area proportional to the observed surface brightness.
\label{Fig2}}
\end{figure}

We have plotted the diffuse flux in one of the {\it FUSE} bands (1B1 at 1117 \AA) as circles with areas proportional to the observed surface brightness in Fig. \ref{Fig2}. There is an excellent correlation between the diffuse FUV flux observed in 1B1 and each of the other bands with linear correlation coefficients of better than 0.9 in each case and a similar plot would be obtained for any of the other bands. The intensities range from around $10^{3}$ photons cm$^{-2}$ s$^{-1}$ sr$^{-1}$ \AA$^{-1}$  to as high as $3 \times 10^{5}$ photons cm$^{-2}$ s$^{-1}$ sr$^{-1}$ \AA$^{-1}$ near 30 Doradus. Although we have listed all the diffuse values for completeness, we note that the effective lower limit for the detection of diffuse radiation with FUSE is about 2000 photons cm$^{-2}$ s$^{-1}$ sr$^{-1}$ \AA$^{-1}$ \citep{Murthy04}. The LMC is at a high Galactic latitude where the contribution from the Galactic diffuse radiation will be relatively small and we have estimated it to be on the order of 500 photons cm$^{-2}$ s$^{-1}$ sr$^{-1}$ \AA$^{-1}$, based on nearby Voyager observations by \citet{Murthy99}.

\citet{Cole99a} found that their diffuse light (2150 \AA) was dominated by the N11 complex in the northwest LMC with a surface brightness of 10$^{4}$ photons cm$^{-2}$ s$^{-1}$ sr$^{-1}$ \AA$^{-1}$, which they identified as a giant reflection nebula. Although, certainly bright in our {\it FUSE} observations, by far the brightest of our observed regions is around 30 Doradus (the Tarantula Nebula), which was not observed by {\it WISP}. As mentioned above, the {\it FUSE} targets have been selected for their proximity to bright stars and, as a result, almost all of our observed diffuse regions are also bright. The FUV diffuse emission in the LMC is predominantly due to the scattering of the star light from the OB associations and shows a large variation in a small angular scale, particularly in 30 Doradus and N11 regions (Fig. \ref{Fig2}). These regions are rich in hot OB stars of UV magnitudes less than 10 \citep{Parker98} and show significant variation in stellar density in a small angular scale.

\begin{figure}
\plotone{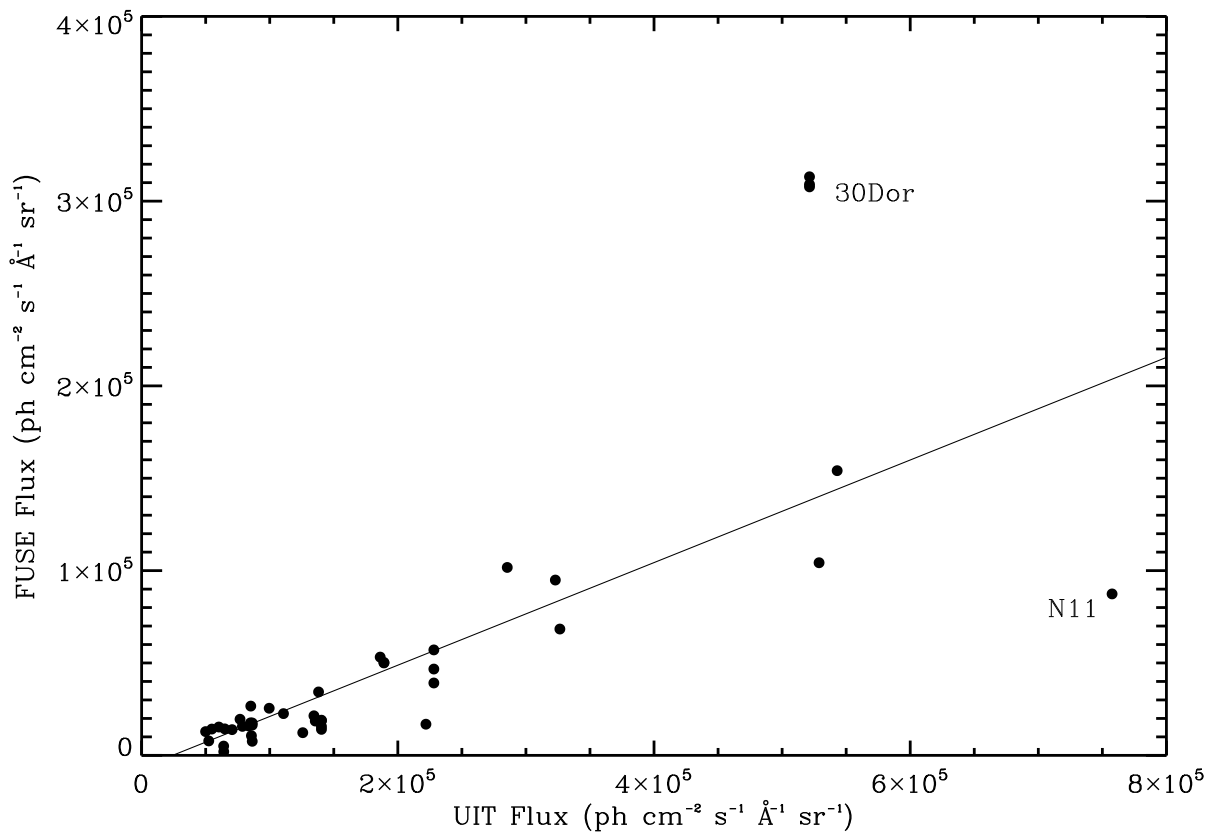}
\caption{Correlation between {\it FUSE} (1B1) and {\it UIT} surface brightness is shown. The correlation coefficient is 0.78 but rises to 0.92 if 4 points (three of the 30Dor and one of the N11 points) are removed. The best fit is the line with slope 0.28 and an offset of -6730 cm$^{-2}$ s$^{-1}$ sr$^{-1}$ \AA$^{-1}$. Similar plots are obtained for the other 6 FUSE bands. Errors in the observations (Table \ref{tbl-1}) are small (relative to Y-axis scale) to be visible and are not shown.   
\label{Fig3}}
\end{figure}

Of the 81 {\it FUSE} locations, 43 overlapped with 10 {\it UIT} field of observations in the LMC and we have found a strong correlation between {\it FUSE} and {\it UIT} diffuse flux (Fig. \ref{Fig3}), where the UIT fluxes have been integrated over the \(30''\) x \(30''\) FUSE aperture. Three points in the 30 Doradus region have a relatively higher FUSE flux because the radiation field is dominated by O stars in the nebula whereas one observation in N11 had a relatively higher UIT flux because the stellar spectrum is much flatter than in other regions. Rather than deal with absolute values, a useful comparison is to find the fractional amount of diffuse radiation in the field defined as diffuse radiation over total radiation (diffuse + stellar) in each of the UIT regions, as was done by \citet{Parker98}. We used the stars from his catalog, translated them into the FUSE spectral range assuming Kurucz models \citep{Kurucz92} and summed the fluxes in a given field. With the observed FUSE/UIT diffuse ratio, we could then calculate the total amount of diffuse radiation in each of the FUSE bands and thus the fraction of total light emitted as diffuse emission. These fractions range from 5\% to 20\% of the total at 1100 \AA, with a high of 45\% in the superbubble N70, with an observed error of 12 -- 17\%. The comparable estimate for the Milky Way is 10\% \citep{Parravano03}.

\begin{figure}
\plotone{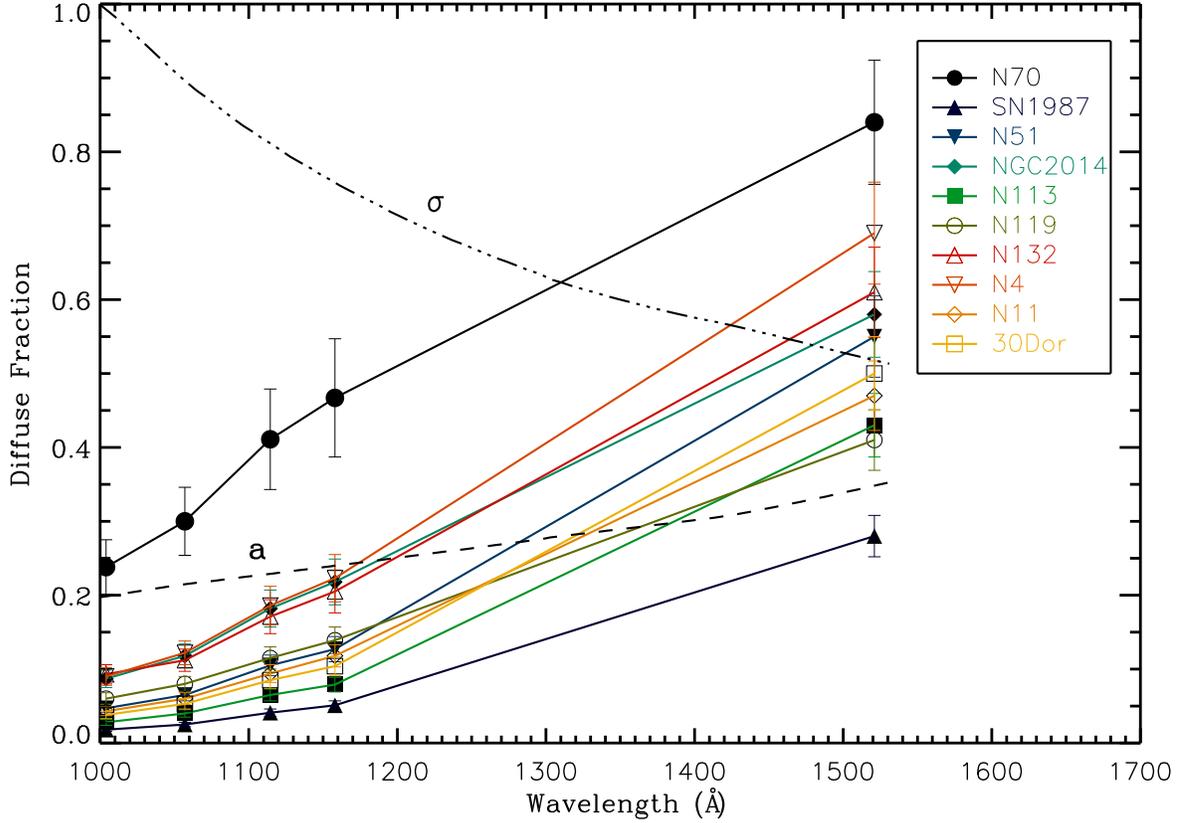}
\caption{Variation of diffuse fraction against {\it FUSE} and {\it UIT} wavelength bands for different regions of LMC. Also plotted are the albedo (dash line) and cross-section (dot-dash line divided by $7.6 \times 10^{-22})$ from model calculations by \citet{Weingartner01}. The seven observed {\it FUSE} bands are shown at four wavelengths (1004, 1057, 1114.5 and 1158 \AA), where the intensities at 1057, 1114.5 and 1158 \AA~ are average of the 1A2 \& 2B1 bands, 1B1 \& 2A2 bands, and 1B2 \& 2A1 bands respectively. The error bars were empirically calculated by taking the extremes of the observed fluxes and range from 12 - 17\% of the data.
\label{Fig4}}
\end{figure}

Although some part of the variation of diffuse fraction in different regions of the LMC may be due to the dust distribution, it is likely that much of the stellar radiation is non-local, as noted by \citet{Cole99a}; i.e., the diffuse light may come from stars far away from the observed area. In our Galaxy, this is seen as scattering of galactic plane star light by high latitude dust clouds \citep{Jura80}; in the LMC, light from the OB associations will be scattered by distant dust. The shape of the diffuse fraction (diffuse radiation over total radiation) is essentially the same in all the 10 regions (Fig. \ref{Fig4}) rising by a factor of about 5 from 1000 to 1500 \AA, implying that most of the heating of the interstellar dust comes in the FUV. This is consistent with the lower albedo (dashed line) and higher cross-section (dot-dash line) of the grains in the FUV. 

\citet{Cole99b} have attempted to model the distribution of diffuse light in the {\it WISP} data by scattering the light of OB associations in the LMC from a dust distribution which decays exponentially with radius and with a hyperbolic secant with distance from the plane. They found that although their models did match the overall morphology of the observations, it was difficult to constrain the parameters because of the uncertainty in many of the physical properties of the ISM in the LMC, particularly in its clumping. Nevertheless they did find that 30 Doradus dominated the diffuse emission in the eastern LMC, a conclusion borne out by our {\it FUSE} observations in the FUV.

\section{Conclusions}

We have obtained the first FUV (1000 -- 1150 \AA) spectra of the diffuse radiation in an external galaxy using serendipitous {\it FUSE} observations of targets in the LMC. Most of these observations are near OB associations and the diffuse emission is bright, ranging from 5\% to 20\% of the total flux in the region. This fraction is much less than the corresponding fraction emitted at 1500 \AA~ suggesting that the largest part of the heating of the interstellar dust occurs in the FUV.

We are now building a more detailed model to use the wealth of data now available in the LMC, particularly the data from {\it Spitzer} \citep{Meixner06}. The UV and IR data are complementary in that they probe different aspects of the radiative transfer between stars and the dust with the part of the radiation not scattered in the UV radiated in the IR and an understanding of the absorption and subsequent re-emission of the starlight in a nearby galaxy such as the LMC will provide templates for more distant galaxies where only the convolution of the two is seen.

\section{Acknowledgements}
This research has made use of {\it Far Ultraviolet Spectroscopic Explorer} data. {\it FUSE} was operated by Johns Hopkins University for NASA. We acknowledge the use of NASA Astrophysics Data System (ADS).

AP is supported by a Department of Science and Technology (DST) research grant. Part of this work was supported through funding from ISRO through the Office of Space Science.

\newpage
\begin{deluxetable}{ccccccccccc}
\tabletypesize{\scriptsize}
\rotate
\tablecaption{Details of {\it FUSE} observations \label{tbl-1}}
\tablewidth{0pt}
\tablehead{
\colhead{Target Name} & \colhead{RA (LWRS)\tablenotemark{a}} & \colhead{Dec (LWRS)\tablenotemark{a}} & \colhead{LiF 1A1\tablenotemark{b}} & \colhead{LiF 1A2\tablenotemark{b}} & \colhead{LiF 1B1\tablenotemark{b}} & \colhead{LiF 1B2\tablenotemark{b}} & \colhead{LiF 2A1\tablenotemark{b}} & \colhead{LiF 2A2\tablenotemark{b}} & \colhead{LiF 2B1\tablenotemark{b}} & \colhead{{\it UIT}\tablenotemark{c}}
}
\startdata
SNR0449-693 & 04 49 40.80 & -69 21 36.0 & 0.74 $\pm$ 0.39 & 0.63 $\pm$ 0.24 & 1.01 $\pm$ 0.14 & 0.91 $\pm$ 0.15 & 0.50 $\pm$ 0.11 & 0.69 $\pm$ 0.12 & 2.06 $\pm$ 0.62 & \\ 
SNR0450-709 & 04 50 28.80 & -70 50 24.0 & 0.36 $\pm$ 0.22 & 0.34 $\pm$ 0.13 & 0.49 $\pm$ 0.10 & 0.69 $\pm$ 0.15 & 0.17 $\pm$ 0.10 & 0.08 $\pm$ 0.03 & 1.11 $\pm$ 0.47 & \\ 
SK-67005 & 04 50 36.00 & -67 38 24.0 & 1.05 $\pm$ 0.60 & 0.52 $\pm$ 0.36 & 0.63 $\pm$ 0.20 & 0.43 $\pm$ 0.17 & 0.63 $\pm$ 0.20 & 0.43 $\pm$ 0.17 & 0.63 $\pm$ 0.20 & \\ 
SNR0454-672 & 04 54 33.60 & -67 12 36.0 & 0.57 $\pm$ 0.13 & 1.12 $\pm$ 0.08 & 1.23 $\pm$ 0.12 & 1.33 $\pm$ 0.10 & 0.80 $\pm$ 0.1 & 0.57 $\pm$ 0.08 & 1.13 $\pm$ 0.13 & 12.58\\ 
SNR0454-665 & 04 54 48.00 & -66 25 48.0 & 1.19 $\pm$ 0.30 & 2.54 $\pm$ 0.18 & 1.40 $\pm$ 0.23 & 1.52 $\pm$ 0.25 & 0.93 $\pm$ 0.17 & 0.91 $\pm$ 0.27 & 1.42 $\pm$ 0.24 & 7.06\\
SK-67D14-BKGD & 04 54 52.80 & -67 15 00.0 & 2.78 $\pm$ 0.12 & 2.87 $\pm$ 0.07 & 2.55 $\pm$ 0.08 & 2.70 $\pm$ 0.11 & 3.13 $\pm$ 0.09 & 2.90 $\pm$ 0.07 & 3.14 $\pm$ 0.05 & 9.96\\ 
SNR0455-687 & 04 55 43.20 & -68 39 00.0 & 0.76 $\pm$ 0.44 & 0.57 $\pm$ 0.27 & 0.63 $\pm$ 0.13 & 0.65 $\pm$ 0.15 & 0.21 $\pm$ 0.09 & 0.15 $\pm$ 0.07 & 1.12 $\pm$ 0.60 & \\ 
LH103204 & 04 56 43.20 & -66 21 36.0 & 1.07 $\pm$ 0.17 & 1.28 $\pm$ 0.11 & 1.95 $\pm$ 0.17 & 1.89 $\pm$ 0.18 & 1.41 $\pm$ 0.16 & 1.37 $\pm$ 0.14 & 1.28 $\pm$ 0.20 & 7.68\\
PGMW-3223 & 04 57 09.60 & -66 21 00.0 & 0.89 $\pm$ 0.25 & 1.26 $\pm$ 0.15 & 1.76 $\pm$ 0.26 & 1.92 $\pm$ 0.36 & 2.29 $\pm$ 0.16 & 0.67 $\pm$ 0.11 & 1.80 $\pm$ 0.22 & 8.49\\
SK-68D15 & 04 57 09.60 & -68 20 24.0 & 0.54 $\pm$ 0.24 & 0.66 $\pm$ 0.14 & 0.92 $\pm$ 0.13 & 1.15 $\pm$ 0.19 & 0.57 $\pm$ 0.15 & 0.49 $\pm$ 0.11 & 0.85 $\pm$ 0.25 & \\ 
LH103073 & 04 57 12.00 & -66 22 48.0 & 6.03 $\pm$ 0.13 & 8.21 $\pm$ 0.08 & 10.17 $\pm$ 0.11 & 9.72 $\pm$ 0.15 & 9.27 $\pm$ 0.44 & 8.08 $\pm$ 0.28 & 6.46 $\pm$ 0.37 & 28.5\\
PGMW-3053 & 04 57 12.00 & -66 23 24.0 & 7.63 $\pm$ 0.15 & 9.95 $\pm$ 0.09 & 10.43 $\pm$ 0.17 & 12.76 $\pm$ 0.19 & 12.92 $\pm$ 0.13 & 11.57 $\pm$ 0.12 & 8.83 $\pm$ 0.12 & 52.88\\ 
PGMW-3070 & 04 57 14.40 & -66 23 24.0 & 6.46 $\pm$ 0.14 & 7.98 $\pm$ 0.09 & 7.87 $\pm$ 0.17 & 8.64 $\pm$ 0.20 & 10.78 $\pm$ 0.17 & 10.11 $\pm$ 0.16 & 8.74 $\pm$ 0.16 & 75.75\\  
SK-68D16 & 04 57 19.20 & -68 21 36.0 & 1.61 $\pm$ 0.91 & 0.92 $\pm$ 0.56 & 1.38 $\pm$ 0.26 & 1.59 $\pm$ 0.33 & 1.38 $\pm$ 0.29 & 0.86 $\pm$ 0.26 & 1.52 $\pm$ 0.93 & \\
PGMW-3168 & 04 57 26.40 & -66 22 48.0 & 1.95 $\pm$ 0.11 & 2.48 $\pm$ 0.07 & 2.67 $\pm$ 0.11 & 2.83 $\pm$ 0.10 & 3.52 $\pm$ 0.12 & 3.35 $\pm$ 0.13 & 2.66 $\pm$ 0.13 & 8.52\\
\enddata
\tablecomments{
Table \ref{tbl-1} is published in its entirety in the 
electronic edition of the {\it Astrophysical Journal Letters}. A portion is 
shown here for guidance regarding its form and content.}
\tablenotetext{a}{Units of right ascension are hours, minutes, and seconds; units of declination are in degrees, arc minutes, and arc seconds.}
\tablenotetext{b}{The surface brightness of the diffuse radiation observed in the {\it FUSE} bands are in units of 10$^{4}$ photon cm$^{-2}$ s$^{-1}$ sr$^{-1}$ \AA$^{-1}$ and the uncertainties are 1$\sigma$ error bar.}
\tablenotetext{c}{{\it UIT} surface brightness in units of 10$^{4}$ photon cm$^{-2}$ s$^{-1}$ sr$^{-1}$ \AA$^{-1}$ and the error in the data is around 10\% \citep{Parker98}.}

\end{deluxetable}
\newpage

\newpage

\newpage

\newpage

\end{document}